\documentclass[superscriptaddress,reprint,prb,floatfix]{revtex4-2}
\usepackage{graphicx,color,dcolumn,bm}
\usepackage{amsmath}
\usepackage[colorlinks=true,breaklinks=true,allcolors=blue]{hyperref}

\begin{document}
\title{Band structure of molybdenum disulfide: from first principle to analytical band model}
\author{Cheng-Hsien Yang}
\affiliation{Department of Electrical Engineering, National Chung Hsing University, Taichung 40227, Taiwan}
\author{Yun-Fang Chung}
\affiliation{Department of Electrical Engineering, National Chung Hsing University, Taichung 40227, Taiwan}
\author{Yen-Shuo Su}
\affiliation{Department of Electrical Engineering, National Chung Hsing University, Taichung 40227, Taiwan}
\author{Kuan-Ting Chen}
\affiliation{Department of Electrical Engineering, National Chung Hsing University, Taichung 40227, Taiwan}
\author{Yi-Sheng Huang}
\affiliation{Department of Physics, National Chung Hsing University, Taichung 40227, Taiwan}
\author{Shu-Tong Chang}
\email{stchang@dragon.nchu.edu.tw}
\affiliation{Department of Electrical Engineering, National Chung Hsing University, Taichung 40227, Taiwan}
\date{\today}

\begin{abstract}
A simple band model such as the effective mass approximation (EMA) can be used to quickly obtain the lower-energy region for the band structure of monolayer molybdenum disulfide. But the EMA band model cannot give the correct description for the band structure in the higher-energy region. To address this major issue, we propose an analytical band calculation (ABC) model to study monolayer molybdenum disulfide. Important parameters of the ABC model are obtained by fitting the three-direction band structure of monolayer molybdenum disulfide obtained from the first-principles (FP) method. The proposed ABC model fits well with the FP band structure calculation result for monolayer molybdenum disulfide. We also use the ABC model to calculate physical quantities used in carrier transport such as density of states and group velocity. Our ABC model can be extended and further utilized for calculating the key physical quantities of ballistic transport of 2D semiconductor materials.
\end{abstract}

\maketitle

\section{Introduction}\label{sec:1}
The scaling technique for complementary metal--oxide--semiconductor (CMOS) field-effect transistor devices encounters a bottleneck when the technology node is less than 2 nm. Two different approaches were developed to deal with this. One proposed to produce transistors with nanowires, which could ensure switch control of the grid over the conductive channel. The other continued the original structure but made semiconductors in metal--oxide--semiconductor field-effect transistors (MOSFETs) in extremely thin semiconductor layers, i.e., ultra-thin-body MOSFETs. A semiconductor layer is about the thickness of an atomic layer for two-dimensional (2D) materials. Two-dimensional materials have gained attention for their potential application in the transistor manufacturing process for the future. The International Roadmap for Devices and Systems (IRDS) 2020 predicted that 2D materials will become the optimal option for channel material technology inflection and 2D device applications beyond CMOS by 2028~\cite{r-01}.

Two-dimensional materials, referring to the thickness of the atomic layer, are layered materials formed by the interaction of atoms in each layer. Van der Waals forces act between layers. In comparison with covalent bonds between molecules, van der Waals forces are extremely small, and layers are easily peeled off, which could help in acquiring few-layer or monolayer 2D materials with mechanical techniques. When electrons move among few-layer materials, the limits are different from those on bulk materials. Some physical characteristics of 2D materials are distinct from known bulk materials~\cite{r-02}. Common 2D materials include graphene, silicene, hexagonal boron nitride, transition metal dichalcogenides (TMD) ~\cite{r-02}, and black phosphorus (BP)~\cite{r-03,r-04} with semiconductor properties. The chemical form of TMD materials is MX$_2$, where M is molybdenum (Mo) or tungsten (W) or some other transition metal, and X is sulfur (S), selenium (Se), or tellurium (Te).

Recently, the highest on-current 2D n-type FET using monolayer MoS$_2$ as the channel material was reported~\cite{r-05,r-06}. Therefore, we focus on 2D TMD semiconducting materials such as MoS$_2$ and study new compact band models for these 2D TMD materials as the major research object in this work. There are many theoretical studies of 2D TMDs reported in the literature~\cite{r-07,r-08,r-09,r-10}. Let us briefly review the typical 2D TMD materials. We consider monolayer molybdenum disulfide that shows a hexagonal structure with close packing. A layer of molybdenum disulfide is the ABA form packed with molybdenum atoms and sulfur atoms, with a thickness of approximately 3.19~\AA{}~\cite{r-10}. Unit cells of molybdenum disulfide are like two linked triangular pyramid forms, with sulfur atoms on the upper and lower layers showing one sixth of unit cells and the center being a molybdenum atom. The corresponding first Brillouin zone (BZ) of a unit cell is hexagonal. Several important symmetry points contain $M$, $K$, and $K'$, where $K$ and $K'$ appear at symmetric positions. Two directions are defined; from the top view, it is the armchair direction along the x direction, while the zigzag direction is along the y direction.

There are three theoretical works~\cite{r-10,r-11,r-12} that not only provide valuable technical/detailed calculations on density functional theory (DFT) methods but also introduce and investigate a wide variety of 2D van der Waals (vdW) materials suitable for scaling in the electronics industry. Let us briefly review these for the reader. Recently, Osanloo et al. studied the dielectric properties of many exfoliable vdW materials using first-principles (FP) methods~\cite{r-10}. They calculated the bandgap and electron affinity and estimated the leakage current through the candidate dielectrics. They also discovered six monolayer dielectrics that promise to outperform bulk HfO$_2$. In 2021, Knobloch et al. demonstrated that, even in the most optimistic case, hBN is unlikely a good choice for a gate insulator in nanoscale 2D CMOS logic by theoretical calculations based on DFT and nonequilibrium Green's functions~\cite{r-11}. Li et al. pointed out that DFT with the screened hybrid functional of Heyd, Scuseria, and Ernzerhof (HSE) is better than the exchange-correlation functional of generalized gradient approximation (GGA) to predict bandgaps with an appreciable accuracy and thus allows the screening of various classes of transition-metal-based compounds such as MoS$_2$ at modest computational cost~\cite{r-12}.

For realizing a tight-binding band model of a TMD such as MoS$_2$, Yao et al. studied a three-band tight-binding (TB) model for monolayer group-VIB TMD~\cite{r-13}. They calculated various TMD materials with two different types of TB models~\cite{r-13}. The difference between the first nearest-neighbor (FNN) and the third nearest-neighbor (TNN) lay in the function of the third neighboring atom being considered in the TNN so that simulation accuracy was higher than that of the FNN. Yao's TB model can fit the FP band structure well, but it is not easy to apply to technology computer-aided design (TCAD) device simulation and SPICE models. Esseni et al. studied the strain-induced carrier mobility modulation in single-layer MoS$_2$ and calculated the monolayer MoS$_2$ band, scattering rate, and carrier mobility~\cite{r-14}. Prof. Esseni proposed an effective mass approximation (EMA) model considering first-order non-parabolic correction used in carrier mobility calculations. A generic TB model for monolayer, bilayer, and bulk MoS$_2$ was developed by Guo et al. The $sp^3d^5$ TB model was used for calculating bulk materials as well as monolayer and double-layer molybdenum disulfide; the model was also promoted to other TMD materials. As this literature review shows, a multiband TB band model can be used to fit the band structure of the FP method. However, this type of TB band model is not easy to apply for fast calculation of key physical quantities such as the density of states (DOS) of 2D TMD materials.

The band structure calculation of new 2D semiconducting materials can be executed by the FP method. Using information about the type of material and the atomic structure, the FP method helps us obtain band structure information through accurate but relatively complicated calculations. We can use a simple band model such as EMA to quickly obtain the band structure if we only focus on the lower-energy region for the band structure. But the EMA band model cannot give the correct description for band structure of 2D TMD material in the higher-energy region. Therefore, a new compact band model, with simpler calculation steps and smaller computing time, allows the analysis and understanding of the properties of new materials for further application to TCAD device simulation and the SPICE model of 2D devices as pointed out in the IRDS 2020. This is the major goal of this paper. After a brief introduction and background, the research methodology is described in Sect.~\ref{sec:2}. Then, we use the research method mentioned in Sect.~\ref{sec:2} to propose a new compact band model, the analytical band calculation (ABC) model, and apply it to novel 2D TMD materials such as MoS$_2$, as described in Sect.~\ref{sec:3}. Finally, our work is summarized in Sect.~\ref{sec:4}.

\begin{figure*}[t]
\centering
\includegraphics[width=5.5in]{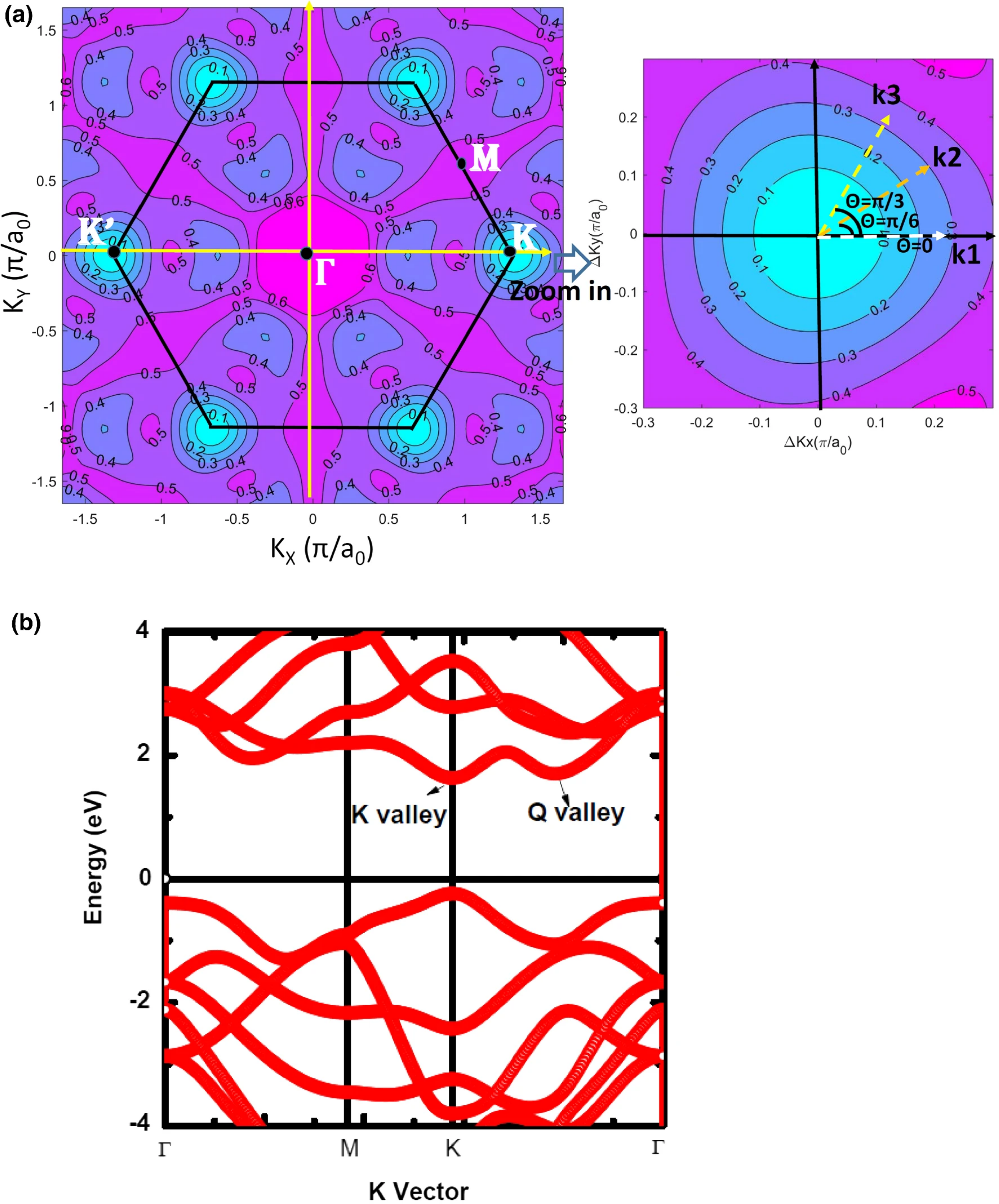}
\caption{(a) The equi-energy contours of monolayer MoS$_2$ in the first BZ. The direct energy gap is located on points $K$ and $K'$, and other symmetry points $\Gamma$ and $M$ are also included in the figure. Zoomed-in equi-energy contours at the $K$ valley are shown in the right-hand side. Note that $\theta=0$, $\pi/6$, $\pi/3$ means the $k$ vector along the $k_1$, $k_2$, $k_3$ directions, respectively. (b) FP band structure of monolayer MoS$_2$. $Q$ valley and $K$ valley are indicated for the reader.}\label{fig:1}
\end{figure*}

\section{Research method}\label{sec:2}
When using the FP method for calculating materials, only the atomic species and lattice structure of materials are required for calculating various physical properties of materials through the corresponding Schrödinger equation. For multi-atom and multi-electron wave function calculation, the Hartree--Fock self-consistent field method and DFT are used for simplifying multi-atom wave functions and various interaction forces. The Vienna Ab initio Simulation Package (VASP) is the one of common FP software packages and was developed by the team of Hafner et al.~\cite{r-16,r-17}. The FP band structure of monolayer MoS$_2$ in this paper was calculated by the VASP~\cite{r-16,r-17} using the projector augmented wave method. The exchange-correlation functional of GGA was used to provide referenced results. The energy cutoff of the plane-wave basis was set to 450~eV, and the convergence criterion was $10^{-6}$~eV. A $\Gamma$-centered $k$ mesh of $18\times18\times1$ was used, and layer separation was greater than $1.8$~nm. For monolayers of MoS$_2$, the lattice constant was optimized, and atomic positions were relaxed until the force on each atom was less than $0.05$~eV/nm. The ABC model, in comparison with the FP method from the VASP, is an easier and less complex calculation method. It shows high accuracy on the fitting with the FP method from the VASP in band structure calculation. In the ABC model for monolayer MoS$_2$, the parameters are acquired by fitting the calculation results of the FP method from the VASP~\cite{r-13,r-17}.

After reviewing the FP method, we introduce a simple band model, namely the effective mass approximation (EMA) model. The EMA model, similar to classical mechanics, calculates the energy and momentum relationship (like the $E$--$K$ relationship for the band structure) for carriers under an external force. Note that effective mass here is not inertial mass, but a proportionality constant to acceleration when electrons experience an external force. The EMA model can be used for band structure calculation around the conduction/valence band edge~\cite{r-08}. The relation for an ideal 2D TMD material is given as
\begin{equation}\label{eq:1}
    E(k)=\frac{\hbar^2k_x^2}{2m_x^*}+\frac{\hbar^2k_y^2}{2m_y^*}=\frac{\hbar^2 k^2}{2m}
\end{equation}

Note that $m_x^*=m_y^*=m$ when no anisotropic effect is considered for the EMA band model. Dr. Kaasbjerg used this EMA model to calculate the scattering rate and applied it to electron mobility. Thus, we call this EMA model Kaasbjerg's model. To enhance the accuracy of the EMA model to cope with a complicated band structure, the EMA model can be used with non-parabolic modification. As the first-order non-parabolic EMA model proposed by Esseni et al., the first-order non-parabolic coefficient $\alpha$ was added to the original EMA model to change the relation to~\cite{r-14}
\begin{equation}\label{eq:2}
    E(1+\alpha E)=\frac{\hbar^2k^2}{2m}
\end{equation}

In this paper, we call this modified EMA model Esseni's band model. Prof. Ridley proposed further modification when calculating the split of the valence band of InAs~\cite{r-18}, where the second-order non-parabolic coefficient $\beta$ was considered. We applied his concept to 2D materials as shown in Eq.\eqref{eq:3}:
\begin{equation}\label{eq:3}
    E(1+\alpha E+\beta E^2)=\frac{\hbar^2k^2}{2m}
\end{equation}

In this paper, we call this second-order non-parabolic EMA model Ridley's band model. With the EMA model, the formula of the physical quantity commonly used in semiconductors can be easily derived as follows:
\begin{equation}\label{eq:4}
    E(k)=\frac{\hbar^2(k_x^2+k_y^2)}{2m}=\frac{\hbar^2k^2}{2m}
\end{equation}

Transforming Eq.\eqref{eq:4} into polar coordinates,
\begin{equation}\label{eq:5}
    k_x^2+k_y^2=k^2,\quad k=\sqrt{\frac{2mE}{\hbar^2}}
\end{equation}

The total status number (N) is given in Eq.\eqref{eq:6}.
\begin{equation}\label{eq:6}
    N=\frac{\iint kdkd\theta}{\left(\frac{2\pi}{L}\right)^2}=\frac{\pi k^2}{\left(\frac{2\pi}{L}\right)^2}
    =\frac{L^2\pi k^2}{4\pi^2}=\frac{L^2}{4\pi}k^2
\end{equation}

The density of state per spin is given in Eq.\eqref{eq:7}.
\begin{equation}\label{eq:7}
    D(E)=\frac{1}{L^2}\frac{dN}{dE}=\frac{1}{L^2}\frac{dN/dk}{dE/dk}=\frac{1}{L^2}\frac{\frac{L^2k}{2\pi}}{\frac{\hbar^2k}{m}}=\frac{m}{2\pi\hbar^2}
\end{equation}

When considering the first-order non-parabolic modified EMA band structure,
\begin{equation}\label{eq:8}
    E(1+\alpha E)=\frac{\hbar^2(k_x^2+k_y^2)}{2m}=\frac{\hbar^2k^2}{2m}
\end{equation}
where
\begin{equation}\label{eq:9}
    k_x^2+k_y^2=k^2,\quad k=\sqrt{\frac{2mE(1+\alpha E)}{\hbar^2}}
\end{equation}

Differentiating k in Eq.\eqref{eq:8},
\begin{equation}\label{eq:10}
    \frac{dE}{dk}(2\alpha E+1)=\frac{\hbar^2k}{m}
\end{equation}

Using the chain rule,
\begin{equation}\label{eq:11}
    \frac{d}{dE}\frac{dE}{dk}(\alpha E^2+E)=\frac{d}{dk}(\alpha E^2+E)=\frac{d}{dk}\frac{\hbar^2k^2}{2m}
\end{equation}

According to the equation of density of state per spin, we get
\begin{equation}\label{eq:12}
    D(E)=\frac{1}{L^2}\frac{\frac{2kL^2}{4\pi}}{\frac{k\hbar^2}{m(1+2\alpha E)}}=\frac{m(1+2\alpha E)}{2\pi\hbar^2}
\end{equation}

In Table~\ref{tab:1}, we summarize EMA, the first- and second-order non-parabolic corrected EMA models, and their analytical physical formulas for density of states per spin as mentioned above.

\begin{table*}[t]
\caption{Three EMA models for monolayer MoS$_2$ and their corresponding analytical physical formula for density of states.}\label{tab:1}
\begin{ruledtabular}
\begin{tabular}{lll}
Type of band model & Analytical formula of band structure & Density of state (per spin)\\  \hline
EMA (Kaasbjerg's model) & $E=\frac{\hbar^2(k_x^2+k_y^2)}{2m^*}$ & $D(E)=\frac{m}{2\pi\hbar^2}$  \\
First-order non-parabolic corrected EMA (Esseni's model) & $E(1+\alpha E)=\frac{\hbar^2(k_x^2+k_y^2)}{2m^*}$ & $D(E)=\frac{m(1+2\alpha E)}{2\pi\hbar^2}$ \\
Second-order non-parabolic corrected EMA (Ridley's model) & $E(1+\alpha E+\beta E^2)=\frac{\hbar^2(k_x^2+k_y^2)}{2m^*}$ & $D(E)=\frac{m(1+2\alpha E+3\beta E^2)}{2\pi\hbar^2}$
\end{tabular}
\end{ruledtabular}
\end{table*}

Finally, we show the compact band model proposed in this work. The ABC model used here is slightly different from common compact band models. Being opposite to the common $E$--$K$ relation, the form of $k(E)$ is used, which could help successive calculations of other key physical quantities such as density of states and equi-energy contours for the conduction band and valence band. Another reason our compact band model is named the ABC model is because it has three major energy-related coefficients $A(E)$, $B(E)$, and $C(E)$ under the general form. The general form of the ABC model for 2D TMD material is
\begin{equation}\label{eq:13}
    k(E,\theta)=A(E)+B(E)\times\cos(n\theta)+C(E)\times\cos(2n\theta)
\end{equation}
Symmetry ($n$): determined by the symmetry of the band structure. When the unit cell or the first Brillouin zone (first BZ) of 2D TMD material shows threefold symmetry, the value of $n$ is 3. Directionality ($\theta$): for determining the direction of $k$ moving toward various symmetry points. Energy effect: determined by $A(E)$, $B(E)$, and $C(E)$ functions. The form of functions $A(E)$, $B(E)$, and $C(E)$ is the combination of $k$ in three different directions. Three boundary conditions in the ABC model are needed to solve $A(E)$, $B(E)$, and $C(E)$ functions. We consider threefold symmetry ($n=3$) as the typical example of 2D TMD material such as monolayer MoS$_2$. The first BZ of molybdenum disulfide as shown in Fig.~\ref{fig:1}(a) is known to have threefold symmetry that the form of the ABC model shows as
\begin{equation}\label{eq:14}
    k(E,\theta)=A(E)+B(E)\times\cos(3\theta)+C(E)\times\cos(6\theta)
\end{equation}

As for the energy $E(\mathbf{k})$, we express the magnitude $k$ versus the energy $E$ along the three symmetry directions $i$ (1, 2, 3) in the BZ. For the monolayer MoS$_2$, such $i$ directions form an angle $\theta=0$, $\pi/6$, and $\pi/3$ with respect to the $\Delta k_x$ direction, which is taken as the transport direction ($k_x$ is armchair direction). Refer to Fig.~\ref{fig:1}(b) for details. The magnitude $k$ of the wave vector along the $i$ direction is expressed as
\begin{equation}\label{eq:15}
    E(1+\alpha_i E+\beta_i E^2)=\frac{\hbar^2(k_i)^2}{2m_i}
\end{equation}
where $\alpha_i$, $\beta_i$, and $m_i$ are the three parameters of the energy model of 2D TMD band set along the $i$ direction. The boundary condition is still the same as for the Ridley' model as shown in Eq.\eqref{eq:3}, such that
\begin{eqnarray}
    \theta=0,\quad k=k_1 \Rightarrow E(1+\alpha_1E+\beta_1E^2)=\frac{\hbar^2k_1^2}{2m_1}\label{eq:16}\\
    \theta=\frac{\pi}{6},\quad k=k_2 \Rightarrow E(1+\alpha_2E+\beta_2E^2)=\frac{\hbar^2k_2^2}{2m_2}\label{eq:17}\\
    \theta=\frac{\pi}{3},\quad k=k_3 \Rightarrow E(1+\alpha_3E+\beta_3E^2)=\frac{\hbar^2k_3^2}{2m_3}\label{eq:18}
\end{eqnarray}

\begin{figure*}[t]
\centering
\includegraphics[width=5.5in]{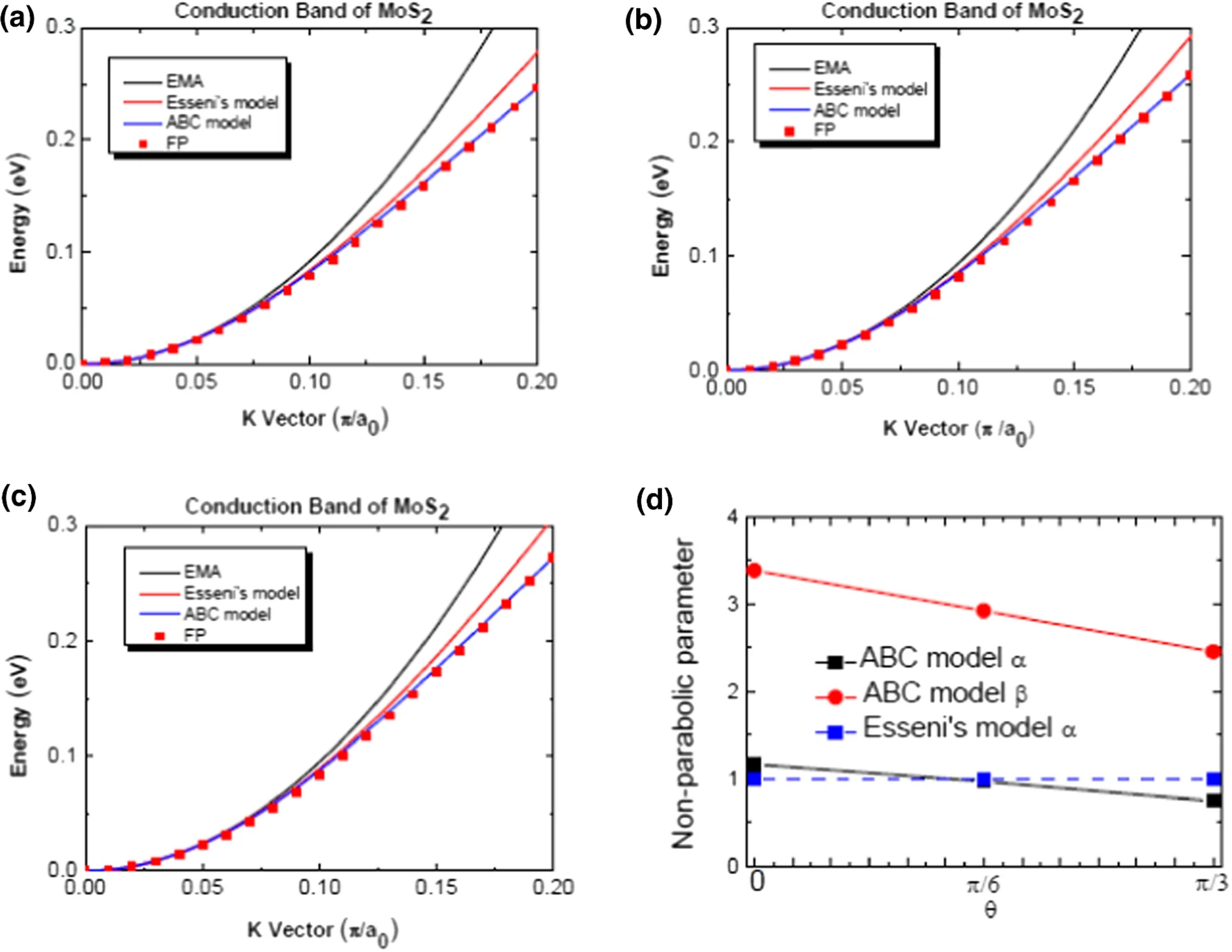}
\caption{Comparison between the conduction band of monolayer MoS$_2$ calculated by the FP method, ABC model, and Esseni's model along the (a) $k_1$ direction, (b) $k_2$ direction, and (c) $k_3$ direction, respectively. Red solid squares in the figure are the calculation results of the FP model, the red line is Esseni's model, and the blue line is the ABC model. EMA is also included for comparison. (d) Non-parabolic parameters used in the ABC model versus the angle $\theta$. The value of $\alpha$ used in Esseni's model~\cite{r-11} is also included for comparison.}\label{fig:2}
\end{figure*}

The form of $k_1$, $k_2$, and $k_3$ can be acquired from the boundary conditions. In the equation, the coefficients $\alpha$, $\beta$, and $m_i$ ($i=1,2,3$) can be acquired by fitting with the FP method. Refer to Fig.~\ref{fig:2} for details. By contrast, when $\theta$ is $0$, $\frac{\pi}{6}$, $\frac{\pi}{3}$, other relations can be acquired as shown in Eqs.\eqref{eq:19}--\eqref{eq:21}, respectively.
\begin{align}
&    k_1=A(E)+B(E)+C(E) \label{eq:19}\\
&    k_2=A(E)-C(E)      \label{eq:20}\\
&    k_3=A(E)-B(E)+C(E) \label{eq:21}
\end{align}

Further solution gives the relationship between key energy-dependent parameters such as $A(E)$, $B(E)$, and $C(E)$ and $k_1$, $k_2$, and $k_3$.

\begin{align}
&    A(E)=\frac{1}{4}[k_1+2k_2+k_3] \label{eq:22}\\
&    B(E)=\frac{1}{2}[k_1-k_3]      \label{eq:23}\\
&    C(E)=\frac{1}{4}[k_1-2k_2+k_3] \label{eq:24}
\end{align}

\begin{figure*}[t]
\centering
\includegraphics[width=5.5in]{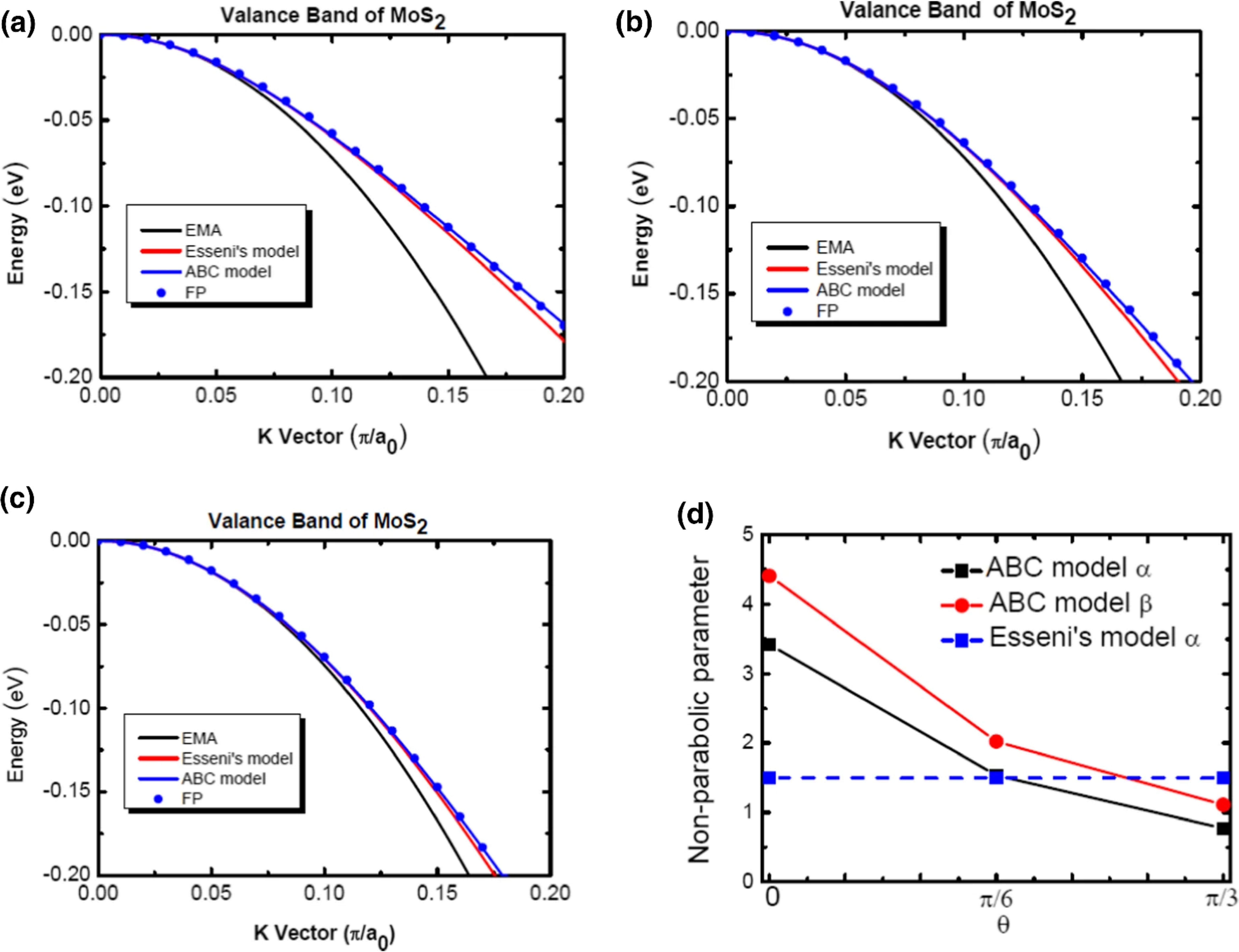}
\caption{Comparison between the valence band of monolayer MoS$_2$ calculated by the FP model, ABC model, and Esseni's model along the (a) $k_1$ direction, (b) $k_2$ direction, and (c) $k_3$ direction, respectively. Blue solid circles in the figure are the calculation results of the FP model, the red line is Esseni's model, and the blue line is the ABC model. EMA is also included for comparison. (d) Non-parabolic parameters used in ABC model versus the angle $\theta$. The value of $\alpha$ used in Esseni's model fitting by this work is also included for comparison.}\label{fig:3}
\end{figure*}

Given the $k$-to-$E$ relation along the $i$ directions, the $k$ versus $\theta$ dependence is finally obtained by noting that for a given $E$ value, $k$ is a periodic function of the angle $\theta$. The period is $2\pi/3$ for the monolayer MoS$_2$. Equations~\eqref{eq:22}--\eqref{eq:24} and \eqref{eq:14} provide the $E(k)$ of the ABC model in an implicit form. Note that Eqs.~\eqref{eq:22}--\eqref{eq:24} can be used for all 2D TMD materials. We summarize parameters used in the ABC model of monolayer MoS$_2$ in Figs.\ref{fig:2}(d) and \ref{fig:3}(d).

Finally, the ABC model is of the form $k(E,\theta)$ and can be applied for calculating the density of states per spin as shown in Eq.\eqref{eq:25}.
\begin{equation}\label{eq:25}
    D(E)=\frac{1}{4\pi^2}\int_0^{2\pi}k(E,\theta)\frac{d}{dE}k(E,\theta)d\theta
\end{equation}

In addition to density of states, there is the averaged group velocity along the x direction, $v_X(E)$, as used in the transport theory of a ballistic nanoscale transistor developed by Prof. Lundstrom's research team from Purdue University~\cite{r-19}. $v_X(E)$ is the average value of $v_x$ over the constant energy surface, which is expressed as Eq.~\eqref{eq:26}:
\begin{equation}\label{eq:26}
    v_x(E)=\frac{\frac{1}{4\pi^2}\int_{\theta=-\pi/2}^{\pi/2} |v_x|\frac{dk}{dE}\kappa d\theta}{\frac{1}{4\pi^2}\int_{\theta=-\pi/2}^{\pi/2} \left(\frac{dk}{dE}\right)kd\theta}
\end{equation}
where the definition of $v_x$ is expressed as $v_X=\frac{1}{\hbar}\frac{\partial E}{\partial k_X}$, and $v_x$ is for the $x$ component of the velocity along the armchair transport direction, so we have
\begin{equation}\label{eq:27}
    v_x=\frac{1}{\hbar}\left[\frac{\partial k}{\partial E}\right]^{-1}\left[\frac{\partial k}{\partial\theta}\frac{\sin\theta}{k}\right]
\end{equation}
which can be analytically expressed as a function of the $E$ and $\theta$. Note that the denominator of Eq.~\eqref{eq:26} is DOS as shown in Eq.~\eqref{eq:25}. We can easily apply Eqs.~\eqref{eq:25}, \eqref{eq:26}, and\eqref{eq:27} to ballistic current calculation for a 2D TMD transistor using the monolayer MoS$_2$ as channel material~\cite{r-19}.

\section{Results and discussion}\label{sec:3}
For the calculations of the band structures, we adopted DFT with the screened hybrid functional of Heyd, Scuseria, and Ernzerhof that has been shown in the literature~\cite{r-20} to produce accurate bandgaps and reasonable effective masses for monolayer MoS$_2$. The HSE functional will give higher bandgap and effective mass than the Perdew--Burke--Ernzerhof (PBE) GGA. Our ABC model can be applied to fit the target band structure from DFT with HSE functional. In this work, no spin-orbit coupling is considered for the band structure calculation of monolayer MoS$_2$. To our knowledge, monolayer MoS$_2$ has a strong spin-orbit coupling (SOC) originated from the $d$ orbitals of the Mo atom, and detail discussion can be found in Ref.\cite{r-17}. The splitting of conduction band minimum (CBM) due to SOC is minimal (${<}10$~meV). With the self-consistent calculations of the spin-orbit effects within the DFT+HSE, Andor Korm\'anyos et al. obtained a split-off value of ${>}100$~meV for the valence band maximum (VBM) at the $K$ point for monolayer MoS$_2$, which is comparable to the experimental data~\cite{r-17}. The spin-orbit effect is an important physical characteristic for monolayer MoS$_2$, but our ABC model does not consider any parameter or function about this effect. Even so, the ABC model can still be easily adopted to fit the target band structure with different spin type. The VASP is one kind of FP calculation software. We use it for calculating the spinless band structure of monolayer molybdenum disulfide to test the validity of the ABC model as proposed in this work. From the band structure, monolayer molybdenum disulfide has a direct energy gap, but is located at point $K$. Several key symmetry points are $K$, $M$, and $K'$, where $\theta$ is the included angle between the plane and $\Delta K_X$-axis, and the BZ shows a hexagon, as illustrated in Fig.~\ref{fig:1}(a). Our calculated FP band structure result for monolayer MoS$_2$ is also plotted in Fig.~\ref{fig:1}(b). We notice that the monolayer MoS$_2$ has a direct bandgap of $1.66$~eV at the $K$ point and an indirect bandgap at the $Q$ point. As you can see in Fig.~\ref{fig:1}(b), there are two bandgaps. For the direct bandgap at the $K$ point, there is an up-shift of the HSE conduction band with respect to the PBE-GGA one, leading to a larger bandgap at the $K$ point (not shown here). For the indirect bandgap at the $Q$ point, the minimum at the $Q$ point is much higher in energy than the minimum at the $K$ point in the case of HSE calculations (not shown here). The main effect from HSE at the $K$ point seems to be that the bandgap increases and the effective masses decrease (not shown here).

\begin{figure}[t]
\includegraphics[width=3.3in]{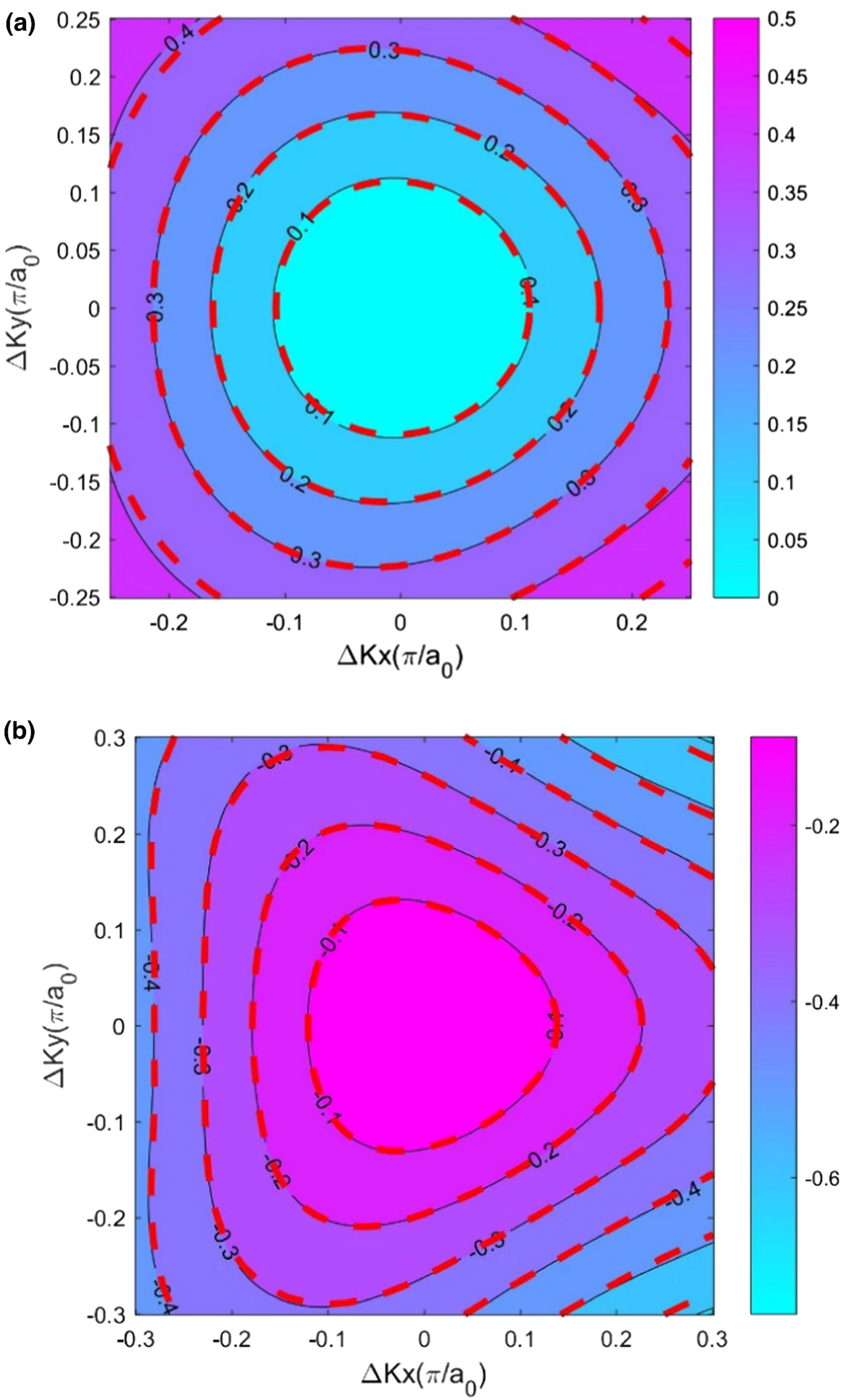}
\caption{Equi-energy contour plot of monolayer MoS$_2$. (a) Equi-energy contour plot around the conduction band minimum. (b) Equi-energy contour plot around the valence band maximum. The background color and black solid lines are the FP model calculation results, and red dotted lines are the ABC model calculation results.}\label{fig:4}
\end{figure}

We use the ABC model to fit the band structure of monolayer MoS$_2$ from the FP method using the VASP. Three $k$ vectors are calculated and compared between the FP method and ABC model. From right to left, they are $k_1$ direction ($\theta=0$), $k_2$ direction ($\theta=\frac{\pi}{6}$), and $k_3$ direction ($\theta=\frac{\pi}{3}$) as shown in Fig.~\ref{fig:1}(b). In the band structure of monolayer MoS$_2$, the ABC model accurately fits the FP method (red circle) in the $k$ value ranging $0$--$0.2$ ($\pi/a_0$). When $k$ is higher than $0.2$ ($\pi/a_0$), the diagram of the band structure appears opposite, which illustrates the difference between Esseni's model and the ABC model. It does not fit well under the Esseni model as shown in Fig.\ref{fig:2} (conduction band) and Fig.\ref{fig:3} (valence band), respectively. Effective masses in the conduction band at the $K$ point along three different angles $\theta$ are $0.401$~$m_0$, $0.394$~$m_0$, $0.390$~$m_0$ in Fig.\ref{fig:2}(a), (b), and (c), respectively. We found that effective masses in the conductor band at the $K$ point are almost unchanged in different angles $\theta$. This implies that the effective mass is isotropic in the $K$ conduction valley. Effective masses in the valence band at the $K$ point along three different angles $\theta$ are $0.514$~$m_0$, $0.513$~$m_0$, and $0.497$~$m_0$ in Fig.\ref{fig:3}(a), (b), and (c), respectively. We found that effective masses in the valence band at the $K$ point are almost unchanged in different angles $\theta$. This implied that the effective mass is also isotropic in the $K$ valence band, similar to the conduction band. Non-parabolic parameters used in the ABC model versus the angle $\theta$ for the conduction band and valence band are shown in Figs.\ref{fig:2}(d) and \ref{fig:3}(d), respectively. As seen in Fig.\ref{fig:2}(d), non-parabolic parameter $\alpha$ from Esseni's model is independent of the angle of $\theta$, while our ABC model is strongly dependent on it. Both non-parabolic parameters $\alpha$ and $\beta$ decrease with increasing angle of $\theta$ from zero to $\pi/3$. Comparing Figs.~\ref{fig:3}(d) and \ref{fig:2}(d), we found that non-parabolic parameters $\alpha$ and $\beta$ are more strongly angle-dependent in the valence band than in the conduction band.

\begin{figure}[t]
\includegraphics[width=3.3in]{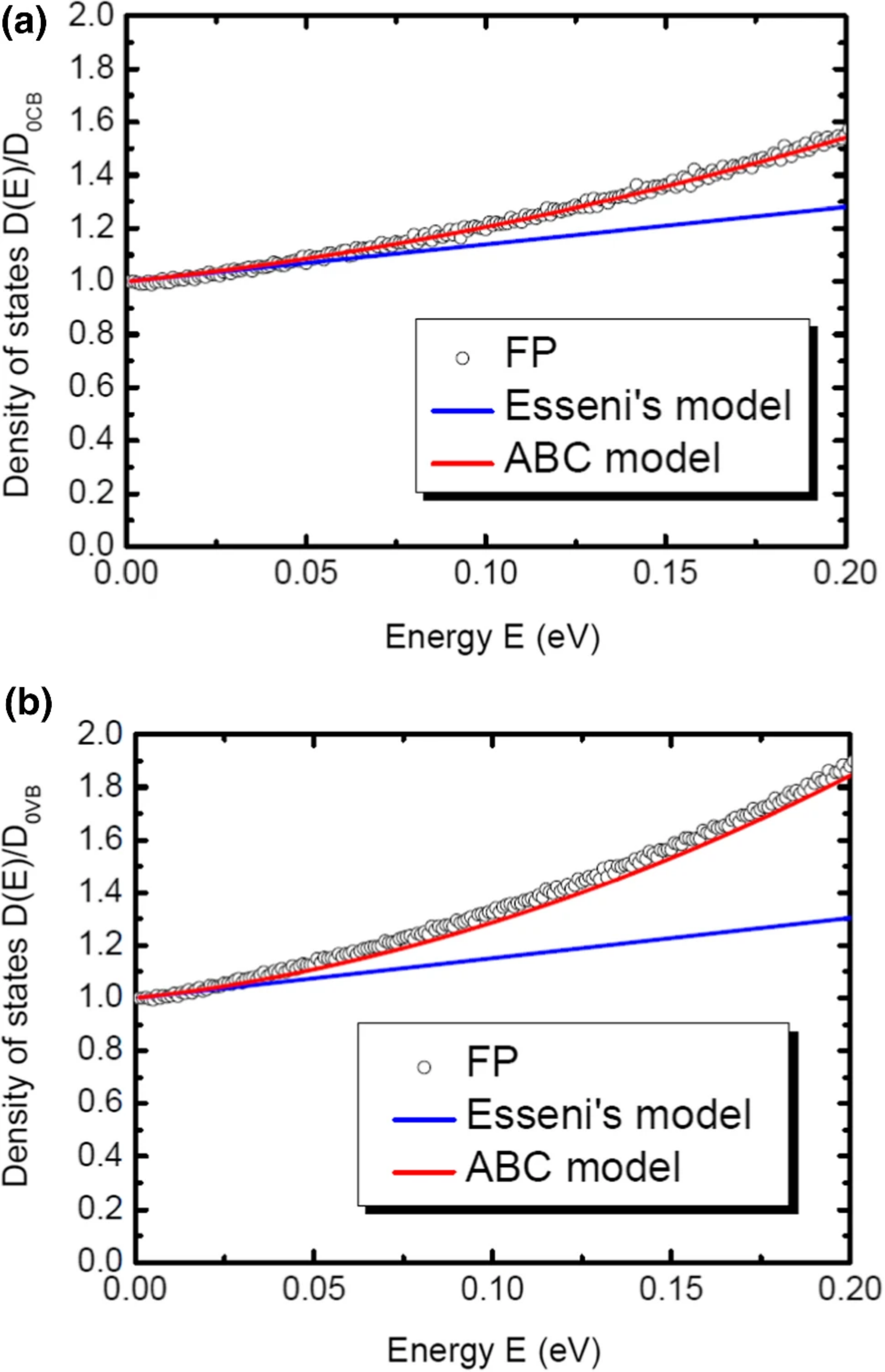}
\caption{Normalized density of states of monolayer MoS$_2$ calculated by three different models, including Esseni's model, the ABC model, and the FP method. (a) Conduction band, density of states $D_{0\mathrm{CB}}$ when energy $E$ is zero. (b) Valence band, density of states $D_{0\mathrm{VB}}$ when energy $E$ equals zero.}\label{fig:5}
\end{figure}

Figure~\ref{fig:4} shows the energy contour plots around the conduction band minimum and the valence band maximum for the monolayer MoS$_2$. With regard to the energy contour plot, the background and black solid lines are the calculation results of the FP model, and the red dotted lines are the results of the ABC model. Apparently, when the energy difference relative to conduction/valence band minima/maxima is smaller than $0.4$~eV, the ABC model fits well with the FP model, regardless of the conduction band or the valence band. As seen in Fig.~\ref{fig:4}(a) and (b), which show the results of FP calculations, in the immediate vicinity of the $K$ point, the dispersion is not isotropic. A trigonal warping of the equi-energy contours can clearly be seen. In comparison to monolayer graphene, we note that its low-energy region is isotropic (energy scale of around $1$~eV), whereas in monolayer MoS$_2$, the trigonal warping is obviously observed at around $100$~meV below the valence band maximum at the $K$ point. We can conclude that trigonal warping is not obvious in the conduction band, but the energy relation clearly presents threefold symmetry in the valence band. It approximately conforms to the calculation results given by Korm\'anyos et al.\cite{r-17}.

\begin{figure}[t]
\includegraphics[width=3.3in]{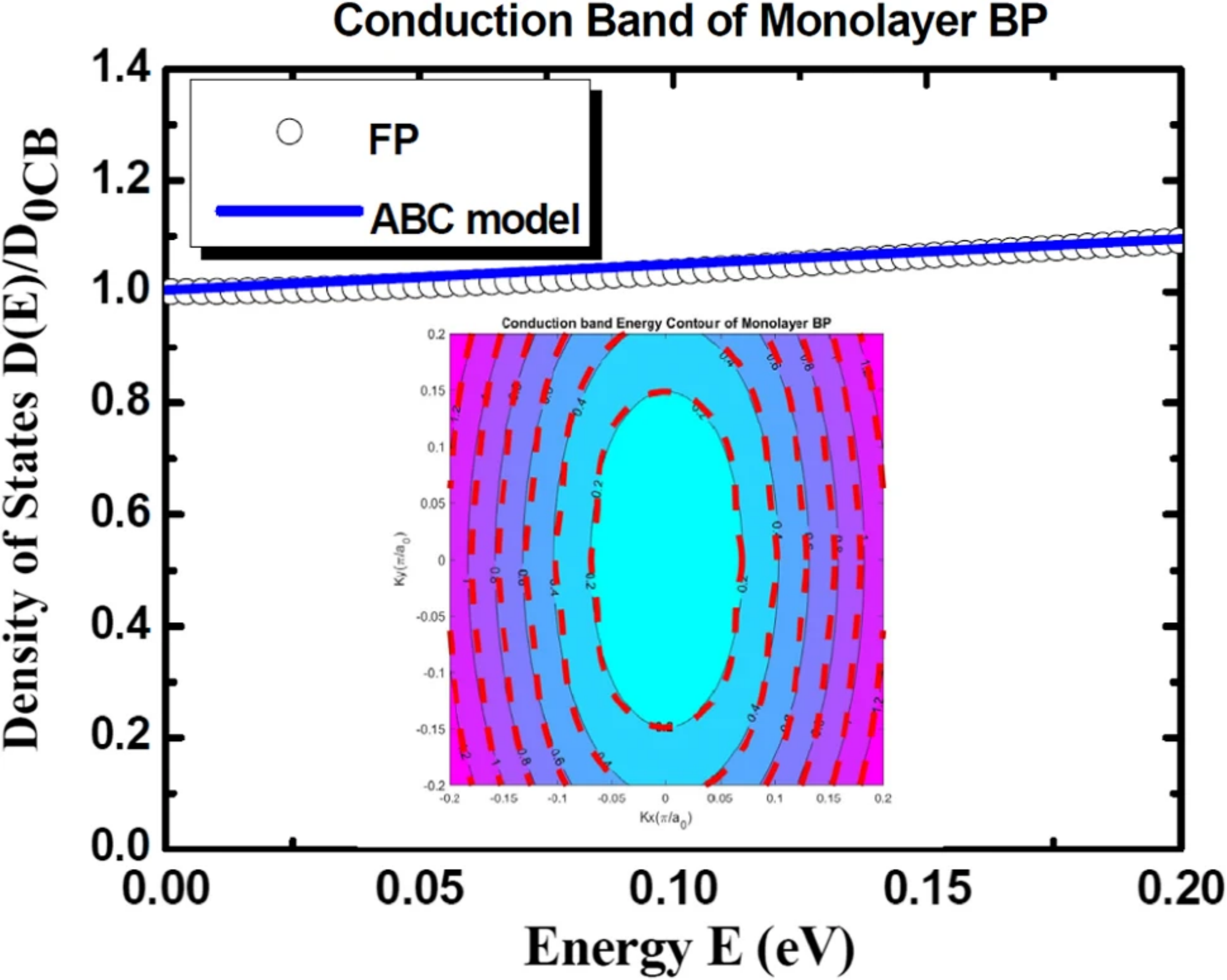}
\caption{Normalized density of states ($D(E)/D_{0\mathrm{CB}}$) of monolayer BP calculated by two different models, including the ABC model and FP method, where $D_{0\mathrm{CB}}$ is defined as density of states at the energy $E$ of zero. Using the averaged non-parabolic parameter $\alpha$ in Esseni's model can fit well with the FP and ABC model. The equi-energy contour plot of BP around the conduction band minimum is also included for the reader's information}\label{fig:6}
\end{figure}

Comparison between of the density of states calculated by the ABC model, Esseni's model, and FP method is shown in Fig.~\ref{fig:5}. Figure~\ref{fig:5}(a) and (b) are for the conduction and valence band, respectively. When energy $=0$, the density of state of the conduction band is defined as $D_{0\mathrm{CB}}$, while it is $D_{0\mathrm{VB}}$ for the valence band, where the value of the valence band is obviously higher than the conduction band owing to larger effective mass. As you can see in Fig.~\ref{fig:5}(a) and (b), which show the results of FP calculation and the ABC model, the density of states is strongly dependent on energy. A quadratic energy dependence on density of states can clearly be seen. In comparison to density of states in the conduction band, we note that the quadratic energy dependence is stronger in the valence band. We also found that Esseni's model can only be correctly fitted in the energy range below $50$~meV. Our ABC model shows good fitting with FP. In Fig.\ref{fig:6}, the ABC model with the Esseni model as the boundary condition (i.e., non-parabolic parameter $\beta\sim0$) can fit the FP band structure of monolayer BP very well. Esseni's model (green line) and the ABC model (blue line) calculation results in the energy interval of $0$--$0.2$~eV appear different for monolayer MoS$2$ as compared to the results from monolayer BP shown in Fig.~\ref{fig:6}; the key coefficients $\alpha$ and $\beta$ show obvious effects. As can be seen in Figs.~\ref{fig:2}(d) and \ref{fig:3}(d), the coefficient $\beta$ of MoS$_2$ is relatively larger than the one of BP (averaged $\alpha<0.5$ and averaged $\beta<0.1$, not shown here). This implies that the ABC model with the Ridley model as the boundary condition is the most suitable band model for monolayer MoS$_2$, while the one with Esseni's model as the boundary condition is good for monolayer BP.

\begin{figure}[t]
\includegraphics[width=3.3in]{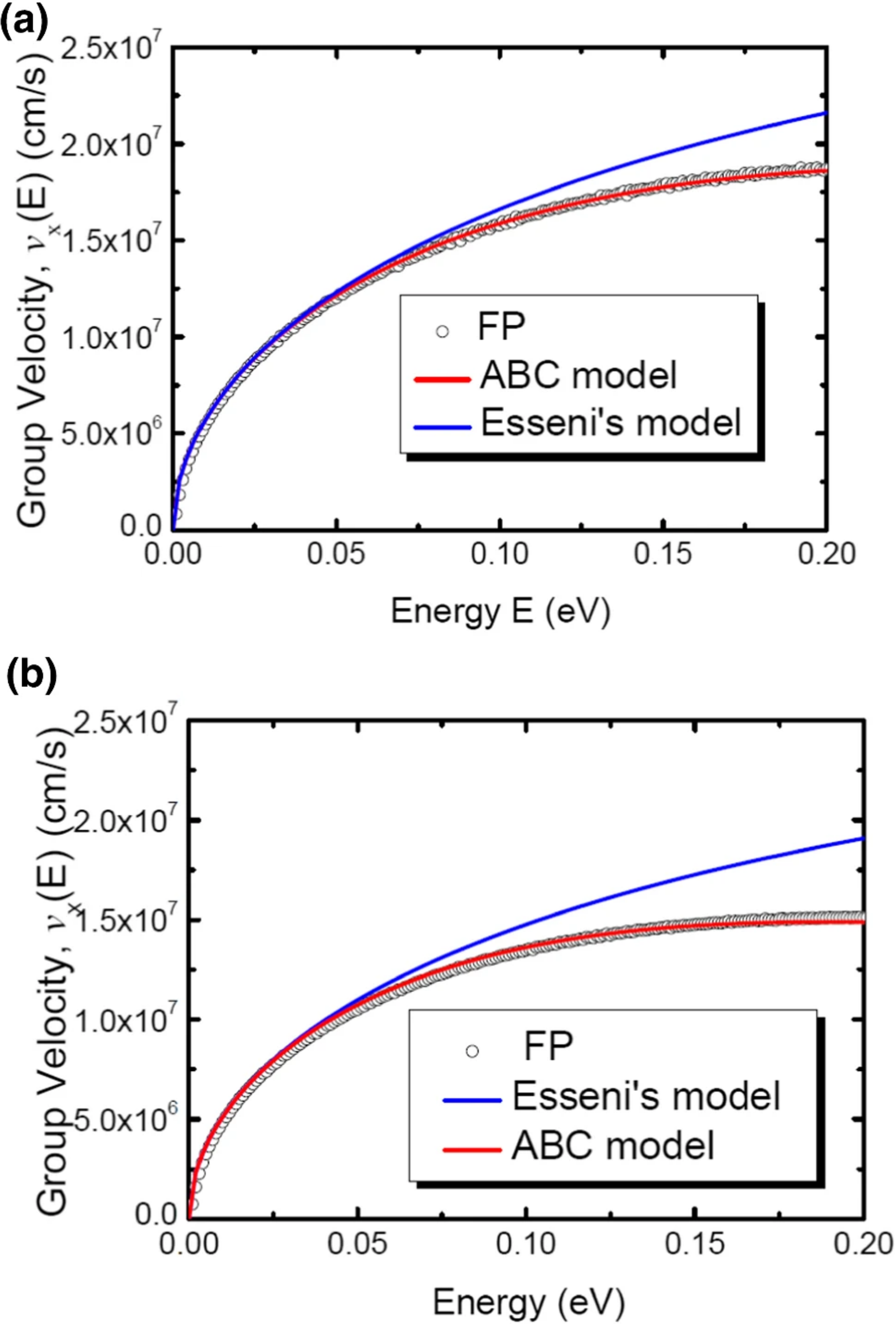}
\caption{Averaged group velocity along the armchair transport direction $v_x(E)$ of monolayer MoS$_2$ as a function of the energy $E$ calculated by three different methods, including the ABC model, Esseni's model, and the FP method. (a) Conduction band. (b) Valence band.}\label{fig:7}
\end{figure}

The comparison of averaged group velocity along the armchair transport direction $v_x(E)$ of monolayer MoS$_2$ either evaluated with the ABC model and Esseni's model or numerically calculated with the FP method is shown in Fig.~\ref{fig:7}. As one can see, our ABC model fits well with the FP method for $v_x(E)$ of monolayer MoS$_2$. We also found that Esseni's model can only be correctly fitted in the energy range below $50$~meV, similar to results of density of states as shown in Fig.~\ref{fig:5}(a) and (b). We also find that the group velocity of the electron is higher than the one of hole owing to smaller effective mass.

The novelty of this work is the use of few parameters by fitting the three-direction band structure obtained from the FP method which can release an analytical band model to describe the band structure in the higher-energy region of monolayer MoS$_2$ and can easily be extended to other 2D semiconductor materials. Our ABC model can save calculation time in the application of mobility and ballistic current calculations.

\section{Conclusion}\label{sec:4}
Our proposed ABC model presents threefold symmetry (MoS$_2$) on band structure calculation, which conforms to the calculation results of the FP model. The ABC model with the Ridley model as the boundary condition can fit the FP band structure of monolayer MoS$_2$ better than that using Esseni's model as the boundary condition. The ABC model can be promoted to $n$-fold symmetry to be suitable for various 2D semiconducting materials. The ABC model is expected to fit quite well with the FP band structure calculation result for various 2D semiconducting materials. Finally, the ABC model can be further utilized for calculating key physical quantities such as carrier mobility and ballistic current of various 2D semiconductor materials.

\begin{acknowledgements}
This work was supported by the National Science Council, Taiwan, R.O.C., under contract nos. MOST110-2622-8-002-014 and MOST 110-2221-E-005-060. Computing support was provided by the National Center for High-Performance Computing (NCHC), Taiwan. We would like to thank Uni-edit (\href{http://www.uni-edit.net}{www.uni-edit.net}) for editing and proofreading this manuscript.
\end{acknowledgements}

%

\end{document}